%Paper: astro-ph/9407002
%From: NESCI@ASTRM1.ASTRO.IT
%Date: Fri, 1 JUL 94 11:08 GMT

\def\spose#1{\hbox to 0pt{#1\hss}}
\def\Dt{\spose{\raise 1.5ex\hbox{\hskip3pt$\mathchar"201$}}}	% upper case
\def\dt{\spose{\raise 1.0ex\hbox{\hskip2pt$\mathchar"201$}}}	% lower case
% This is AA.CMM, the plain TeX macro package
% (CM version) from Springer-Verlag
% for the Astronomy and Astrophysics Main Journal
% Version 2.0 as of 25 Feb 1991
%
% Test for recursive or multiple loading of Springer macro packages
\def\SpringerMacroPackageNameATest{AA}%
\let\next\relax
\ifx\SpringerMacroPackageNameA\undefined
  \message{Loading the \SpringerMacroPackageNameATest\space
           macro package from Springer-Verlag...}%
\else
  \ifx\SpringerMacroPackageNameA\SpringerMacroPackageNameATest
    \message{\SpringerMacroPackageNameA\space macro package
             from Springer-Verlag already loaded.}%
    \let\next\endinput
  \else
    \message{DANGER: \SpringerMacroPackageNameA\space from
             Springer-Verlag already loaded, will try to proceed.}%
  \fi
\fi
\next
\def\SpringerMacroPackageNameA{AA}%
% now call all the sub-macros
% indention of equations
\newskip\mathindent      \mathindent=0pt
% \titlea
\newskip\tabefore \tabefore=20dd plus 10pt minus 5pt      % space above
\newskip\taafter  \taafter=10dd                           % space below
% \titleb
\newskip\tbbeforeback    \tbbeforeback=-20dd   % corrective space to a \titlea
\newskip\tbbefore        \tbbefore=17pt plus 7pt minus3pt % spaceabove
\newskip\tbafter         \tbafter=8pt                     % space below
% \titlec
\newskip\tcbeforeback    \tcbeforeback=-3pt    % corrective space to a \titleb
\advance\tcbeforeback by -10dd                 % corrective space to a \titleb
\newskip\tcbefore        \tcbefore=10dd plus 5pt minus 1pt% space above
\newskip\tcafter         \tcafter=6pt                     % space below
% \titled
\newskip\tdbeforeback    \tdbeforeback=-3pt    % corrective space to a \titlec
\advance\tdbeforeback by -10dd                 % corrective space to a \titlec
\newskip\tdbefore        \tdbefore=10dd plus 4pt minus 1pt   % space above
% \petit
\newskip\petitsurround
\petitsurround=6pt\relax
% \ack
\newskip\ackbefore      \ackbefore=10dd plus 5pt             % space above
\newskip\ackafter       \ackafter=6pt                        % space below
% indention of lists
\newdimen\itemindent    \newdimen\itemitemindent
\itemindent=1.5em       \itemitemindent=2\itemindent
 \font \tatt            = cmbx10 scaled \magstep3
 \font \tats            = cmbx10 scaled \magstep1
 \font \tamt            = cmmib10 scaled \magstep3
 \font \tams            = cmmib10 scaled \magstep1
 \font \tamss           = cmmib10
 \font \tast            = cmsy10 scaled \magstep3
 \font \tass            = cmsy10 scaled \magstep1
 \font \tbtt            = cmbx10 scaled \magstep2
 \font \tbmt            = cmmib10 scaled \magstep2
 \font \tbst            = cmsy10 scaled \magstep2
\catcode`@=11    % use @ as a normal character
\vsize=23.5truecm
\hoffset=-1true cm
\voffset=-1true cm
\normallineskip=1dd
\normallineskiplimit=0dd
\newskip\ttglue%
\def\ifundefin@d#1#2{%
\expandafter\ifx\csname#1#2\endcsname\relax}
\def\getf@nt#1#2#3#4{%
\ifundefin@d{#1}{#2}%
\global\expandafter\font\csname#1#2\endcsname=#3#4%
\fi\relax
}
\newfam\sffam
\newfam\scfam
\def\makesize#1#2#3#4#5#6#7{%
 \getf@nt{rm}{#1}{cmr}{#2}%
 \getf@nt{rm}{#3}{cmr}{#4}%
 \getf@nt{rm}{#5}{cmr}{#6}%
 \getf@nt{mi}{#1}{cmmi}{#2}%
 \getf@nt{mi}{#3}{cmmi}{#4}%
 \getf@nt{mi}{#5}{cmmi}{#6}%
 \getf@nt{sy}{#1}{cmsy}{#2}%
 \getf@nt{sy}{#3}{cmsy}{#4}%
 \getf@nt{sy}{#5}{cmsy}{#6}%
 \skewchar\csname mi#1\endcsname ='177
 \skewchar\csname mi#3\endcsname ='177
 \skewchar\csname mi#5\endcsname ='177
 \skewchar\csname sy#1\endcsname ='60
 \skewchar\csname sy#3\endcsname='60
 \skewchar\csname sy#5\endcsname='60
\expandafter\def\csname#1size\endcsname{%
 \normalbaselineskip=#7
 \normalbaselines
 \setbox\strutbox=\hbox{\vrule height0.75\normalbaselineskip%
    depth0.25\normalbaselineskip width0pt}%
 \textfont0=\csname rm#1\endcsname
 \scriptfont0=\csname rm#3\endcsname
 \scriptscriptfont0=\csname rm#5\endcsname
    \def\oldstyle{\fam1\csname mi#1\endcsname}%
 \textfont1=\csname mi#1\endcsname
 \scriptfont1=\csname mi#3\endcsname
 \scriptscriptfont1=\csname mi#5\endcsname
 \textfont2=\csname sy#1\endcsname
 \scriptfont2=\csname sy#3\endcsname
 \scriptscriptfont2=\csname sy#5\endcsname
 \textfont3=\tenex\scriptfont3=\tenex\scriptscriptfont3=\tenex
   \def\rm{%
 \fam0\csname rm#1\endcsname%
   }%
   \def\it{%
 \getf@nt{it}{#1}{cmti}{#2}%
 \textfont\itfam=\csname it#1\endcsname
 \fam\itfam\csname it#1\endcsname
   }%
   \def\sl{%
 \getf@nt{sl}{#1}{cmsl}{#2}%
 \textfont\slfam=\csname sl#1\endcsname
 \fam\slfam\csname sl#1\endcsname}%
   \def\bf{%
 \getf@nt{bf}{#1}{cmbx}{#2}%
 \getf@nt{bf}{#3}{cmbx}{#4}%
 \getf@nt{bf}{#5}{cmbx}{#6}%
 \textfont\bffam=\csname bf#1\endcsname
 \scriptfont\bffam=\csname bf#3\endcsname
 \scriptscriptfont\bffam=\csname bf#5\endcsname
 \fam\bffam\csname bf#1\endcsname}%
   \def\tt{%
 \getf@nt{tt}{#1}{cmtt}{#2}%
 \textfont\ttfam=\csname tt#1\endcsname
 \fam\ttfam\csname tt#1\endcsname
 \ttglue=.5em plus.25em minus.15em
   }%
  \def\sf{%
\getf@nt{sf}{#1}{cmss}{10 at #2pt}%
\textfont\sffam=\csname sf#1\endcsname
\fam\sffam\csname sf#1\endcsname}%
   \def\sc{%
 \getf@nt{sc}{#1}{cmcsc}{10 at #2pt}%
 \textfont\scfam=\csname sc#1\endcsname
 \fam\scfam\csname sc#1\endcsname}%
\rm }}
\makesize{IXf}{9}{VIf}{6}{Vf}{5}{10.00dd}
\def\normalsize{\IXfsize
\def\sf{%
   \getf@nt{sf}{IXf}{cmss}{9}%
   \getf@nt{sf}{VIf}{cmss}{10 at 6pt}%
   \getf@nt{sf}{Vf}{cmss}{10 at 5pt}%
   \textfont\sffam=\csname sfIXf\endcsname
   \scriptfont\sffam=\csname sfVIf\endcsname
   \scriptscriptfont\sffam=\csname sfVf\endcsname
   \fam\sffam\csname sfIXf\endcsname}%
}%
\newfam\mibfam
\def\mib{%
   \getf@nt{mib}{IXf}{cmmib}{10 at9pt}%
   \getf@nt{mib}{VIf}{cmmib}{10 at6pt}%
   \getf@nt{mib}{Vf}{cmmib}{10 at5pt}%
   \textfont\mibfam=\csname mibIXf\endcsname
   \scriptfont\mibfam=\csname mibVIf\endcsname
   \scriptscriptfont\mibfam=\csname mibVf\endcsname
   \fam\mibfam\csname mibIXf\endcsname}%
\makesize{Xf}{10}{VIf}{6}{Vf}{5}{10.00dd}
\Xfsize
\it\bf\tt\rm

\def\tentt{\ttXf}

\normalsize
\it\bf\tt\sf\mib\rm
\def\boldmath{\textfont1=\mibIXf \scriptfont1=\mibVIf
\scriptscriptfont1=\mibVf}
\newdimen\fullhsize
\newcount\verybad \verybad=1010
\let\lr=L%
\fullhsize=40cc
\hsize=19.5cc
\def\fullline{\hbox to\fullhsize}
\def\makefootline{\baselineskip=10dd \fullline{\the\footline}}
\def\makeheadline{\vbox to 0pt{\vskip-22.5pt
            \fullline{\vbox to 8.5pt{}\the\headline}\vss}\nointerlineskip}
\hfuzz=2pt
\vfuzz=2pt
\tolerance=1000
\abovedisplayskip=3 mm plus6pt minus 4pt
\belowdisplayskip=3 mm plus6pt minus 4pt
\abovedisplayshortskip=0mm plus6pt
\belowdisplayshortskip=2 mm plus4pt minus 4pt
\parindent=1.5em
\newdimen\stdparindent\stdparindent\parindent
\frenchspacing
\nopagenumbers
\predisplaypenalty=600        % Make a page break before a display harder
\displaywidowpenalty=2000     % and even harder for a widow display.
\def\widowsandclubs#1{\global\verybad=#1
   \global\widowpenalty=\the\verybad1      % default: 10101
   \global\clubpenalty=\the\verybad2  }    % default: 10102
\widowsandclubs{1010}
\def\paglay{\headline={{\normalsize\hsize=.75\fullhsize\ifnum\pageno=1
\vbox{\hrule\line{\vrule\kern3pt\vbox{\kern3pt
\hbox{\bf A\&A manuscript no.}
\hbox{(will be inserted by hand later)}
\kern3pt\hrule\kern3pt
\hbox{\bf Your thesaurus codes are:}
\hbox{\rightskip=0pt plus3em\advance\hsize by-7pt
\vbox{\bf\noindent\ignorespaces\the\THESAURUS}}
\kern3pt}\hfil\kern3pt\vrule}\hrule}
\rlap{\quad\AALogo}\hfil
\else\normalsize\ifodd\pageno\hfil\folio\else\folio\hfil\fi\fi}}}
\makesize{VIIIf}{8}{VIf}{6}{Vf}{5}{9.00dd}
      \getf@nt{sf}{VIIIf}{cmss}{8}%
      \getf@nt{sf}{VIf}{cmss}{10 at 6pt}%
      \getf@nt{sf}{Vf}{cmss}{10 at 5pt}%
      \getf@nt{mib}{VIIIf}{cmmib}{10 at 8pt}%
      \getf@nt{mib}{VIf}{cmmib}{10 at 6pt}%
      \getf@nt{mib}{Vf}{cmmib}{10 at 5pt}%
\VIIIfsize\it\bf\tt\rm
\normalsize
\def\petit{\VIIIfsize
   \def\sf{%
      \getf@nt{sf}{VIIIf}{cmss}{8}%
      \getf@nt{sf}{VIf}{cmss}{10 at 6pt}%
      \getf@nt{sf}{Vf}{cmss}{10 at 5pt}%
      \textfont\sffam=\csname sfVIIIf\endcsname
      \scriptfont\sffam=\csname sfVIf\endcsname
      \scriptscriptfont\sffam=\csname sfVf\endcsname
      \fam\sffam\csname sfVIIIf\endcsname
}%
\def\mib{%
   \getf@nt{mib}{VIIIf}{cmmib}{10 at 8pt}%
   \getf@nt{mib}{VIf}{cmmib}{10 at 6pt}%
   \getf@nt{mib}{Vf}{cmmib}{10 at 5pt}%
   \textfont\mibfam=\csname mibVIIIf\endcsname
   \scriptfont\mibfam=\csname mibVIf\endcsname
   \scriptscriptfont\mibfam=\csname mibVf\endcsname
   \fam\mibfam\csname mibIXf\endcsname}%
\def\boldmath{\textfont1=\mibVIIIf\scriptfont1=\mibVIf
\scriptscriptfont1=\mibVf}%
\let\bfIXf=\bfVIIIf
 \if Y\REFEREE \normalbaselineskip=2\normalbaselineskip
 \normallineskip=2\normallineskip\fi
 \setbox\strutbox=\hbox{\vrule height7pt depth2pt width0pt}%
 \normalbaselines\rm}%
\def\begpet{\vskip\petitsurround
\bgroup\petit}%  Beginn eines Paragraphen in petit
\def\endpet{\vskip\petitsurround
\egroup}%  Ende eines Paragraphen in petit

 \let  \tatss           = \bfXf
 \let  \tasss           = \syXf
 \let  \tbts            = \bfXf
 \let  \tbtss           = \bfVIIIf
 \let  \tbms            = \tamss
 \let  \tbmss           = \mibVIIIf
 \let  \tbss            = \syXf
 \let  \tbsss           = \syVIIIf
\def\newline{\hfill\break}% makes a new line in the text :)
\def\rahmen#1{\vbox{\hrule\line{\vrule\vbox to#1true
cm{\vfil}\hfil\vrule}\vfil\hrule}}
\let\ts=\thinspace
\def\,{\relax\ifmmode\mskip\thinmuskip\else\thinspace\fi}
\def\unvskip{%
   \ifvmode
      \ifdim\lastskip=0pt
      \else
         \vskip-\lastskip
      \fi
   \fi}
\newtoks\eq\newtoks\eqn
\newdimen\mathhsize
\def\calcmathhsize{\mathhsize=\hsize
\advance\mathhsize by-\mathindent}
\calcmathhsize
\def\eqalign#1{\null\vcenter{\openup\jot\m@th
  \ialign{\strut\hfil$\displaystyle{##}$&$\displaystyle{{}##}$\hfil
      \crcr#1\crcr}}}
\def\displaylines#1{{}$\displ@y
\hbox{\vbox{\halign{$\@lign\hfil\displaystyle##\hfil$\crcr
    #1\crcr}}}${}}
\def\eqalignno#1{{}$\displ@y
  \hbox{\vbox{\halign
to\mathhsize{\hfil$\@lign\displaystyle{##}$\tabskip\z@skip
    &$\@lign\displaystyle{{}##}$\hfil\tabskip\centering
    &\llap{$\@lign##$}\tabskip\z@skip\crcr
    #1\crcr}}}${}}
\def\leqalignno#1{{}$\displ@y
\hbox{\vbox{\halign
to\mathhsize{\qquad\hfil$\@lign\displaystyle{##}$\tabskip\z@skip
    &$\@lign\displaystyle{{}##}$\hfil\tabskip\centering
    &\kern-\mathhsize\rlap{$\@lign##$}\tabskip\hsize\crcr
    #1\crcr}}}${}}
\def\generaldisplay{%
\ifeqno
       \ifleqno\leftline{$\displaystyle\the\eqn\quad\the\eq$}%
       \else\noindent\kern\mathindent\hbox to\mathhsize{$\displaystyle
             \the\eq\hfill\the\eqn$}%
       \fi
\else
       \kern\mathindent
       \hbox to\mathhsize{$\displaystyle\the\eq$\hss}%
\fi
\global\eq={}\global\eqn={}}%
\newif\ifeqno\newif\ifleqno
\everydisplay{\displaysetup}
\def\displaysetup#1$${\displaytest#1\eqno\eqno\displaytest}
% look for equation numbers
\def\displaytest#1\eqno#2\eqno#3\displaytest{%
\if!#3!\ldisplaytest#1\leqno\leqno\ldisplaytest
\else\eqnotrue\leqnofalse\eqn={#2}\eq={#1}\fi
\generaldisplay$$}
\def\ldisplaytest#1\leqno#2\leqno#3\ldisplaytest{\eq={#1}%
\if!#3!\eqnofalse\else\eqnotrue\leqnotrue\eqn={#2}\fi}
\newcount\eqnum\eqnum=0% register
\def\autnum{\global\advance\eqnum by 1\relax{\rm(\the\eqnum)}}
\newdimen\lindent
\lindent=\stdparindent

\def\litemitem{\par\noindent\hbox to\lindent{\hfil}%
               \hangindent=2\lindent\ltextindent}
\def\ltextindent#1{\hbox to\lindent{#1\hss}\ignorespaces}
\def\set@item@mark#1{\llap{#1\enspace}\ignorespaces}
\ifx\undefined\mathhsize
   \def\item{\par\noindent
   \hangindent\itemindent\hangafter=0
   \set@item@mark}
   \def\itemitem{\par\noindent\advance\mathhsize by-\itemitemindent
   \hangindent\itemitemindent\hangafter=0
   \set@item@mark}
\else
   \def\item{\par\noindent\advance\mathhsize by-\itemindent
   \hangindent\itemindent\hangafter=0
   \everypar={\global\mathhsize=\hsize
   \global\advance\mathhsize by-\mathindent
   \global\everypar={}}\set@item@mark}
   \def\itemitem{\par\noindent\advance\mathhsize by-\itemitemindent
   \hangindent\itemitemindent\hangafter=0
   \everypar={\global\mathhsize=\hsize
   \global\advance\mathhsize by-\mathindent
   \global\everypar={}}\set@item@mark}
\fi
\newcount\the@end \global\the@end=0
\newbox\springer@macro \setbox\springer@macro=\vbox{}
\def\typeset{\setbox\springer@macro=\vbox{\begpet\noindent
   This article was processed by the author using
   Sprin\-ger-Ver\-lag \TeX{} A\&A macro package 1991.\par
   \egroup}\global\the@end=1}
\outer\def\bye{\bigskip\typeset
\sterne=1\ifx\speciali\undefined
\else
  \loop\smallskip\noindent special character No\number\sterne:
    \csname special\romannumeral\sterne\endcsname
    \advance\sterne by 1\relax
    \ifnum\sterne<11\relax
  \repeat
\fi
\if R\lr\null\fi\vfill\supereject\end}
\def\AALogo{\setbox254=\hbox{ ASTROPHYSICS }%
\vbox{\baselineskip=10dd\hrule\hbox{\vrule\vbox{\kern3pt
\hbox to\wd254{\hfil ASTRONOMY\hfil}
\hbox to\wd254{\hfil AND\hfil}\copy254
\hbox to\wd254{\hfil\number\day.\number\month.\number\year\hfil}
\kern3pt}\vrule}\hrule}}
\def\figure#1#2{\medskip\noindent{\petit{\bf Fig.\ts#1.\
}\ignorespaces#2\par}}
\expandafter \newcount \csname c@Tl\endcsname
    \csname c@Tl\endcsname=0
\expandafter \newcount \csname c@Tm\endcsname
    \csname c@Tm\endcsname=0
\expandafter \newcount \csname c@Tn\endcsname
    \csname c@Tn\endcsname=0
\expandafter \newcount \csname c@To\endcsname
    \csname c@To\endcsname=0
\expandafter \newcount \csname c@Tp\endcsname
    \csname c@Tp\endcsname=0
\expandafter \newcount \csname c@fn\endcsname
    \csname c@fn\endcsname=0
\def \stepc#1    {\global
    \expandafter
    \advance
    \csname c@#1\endcsname by 1}
\def \resetcount#1    {\global
    \csname c@#1\endcsname=0}
\def\@nameuse#1{\csname #1\endcsname}
\def\arabic#1{\@arabic{\@nameuse{c@#1}}}
\def\@arabic#1{\ifnum #1>0 \number #1\fi}
 \def \aTa  { \goodbreak
     \bgroup
     \par
 \textfont0=\tatt \scriptfont0=\tats \scriptscriptfont0=\tatss
 \textfont1=\tamt \scriptfont1=\tams \scriptscriptfont1=\tamss
 \textfont2=\tast \scriptfont2=\tass \scriptscriptfont2=\tasss
     \baselineskip=17dd\lineskiplimit=0pt\lineskip=0pt
     \rightskip=0pt plus4cm
     \pretolerance=10000
     \noindent
     \tatt}
 \def \eTa{\vskip10pt\egroup
     \noindent
     \ignorespaces}
 \def \aTb{\goodbreak
     \bgroup
     \par
 \textfont0=\tbtt \scriptfont0=\tbts \scriptscriptfont0=\tbtss
 \textfont1=\tbmt \scriptfont1=\tbms \scriptscriptfont1=\tbmss
 \textfont2=\tbst \scriptfont2=\tbss \scriptscriptfont2=\tbsss
     \baselineskip=13dd\lineskip=0pt\lineskiplimit=0pt
     \rightskip=0pt plus4cm
     \pretolerance=10000
     \noindent
     \tbtt}
 \def \eTb{\vskip10pt
     \egroup
     \noindent
     \ignorespaces}
\newcount\section@penalty  \section@penalty=0
\newcount\subsection@penalty  \subsection@penalty=0
\newcount\subsubsection@penalty  \subsubsection@penalty=0
\def\titlea#1{\par\stepc{Tl}
    \resetcount{Tm}
    \bgroup
       \normalsize
       \bf \rightskip 0pt plus4em
       \pretolerance=20000
       \boldmath
       \setbox0=\vbox{\vskip\tabefore
          \noindent
          \arabic{Tl}.\
          \ignorespaces#1
          \vskip\taafter}
       \dimen0=\ht0\advance\dimen0 by\dp0
       \advance\dimen0 by 2\baselineskip
       \advance\dimen0 by\pagetotal
       \ifdim\dimen0>\pagegoal
          \ifdim\pagetotal>\pagegoal
          \else\eject\fi\fi
       \vskip\tabefore
       \penalty\section@penalty \global\section@penalty=-200
       \global\subsection@penalty=10007
       \noindent
       \arabic{Tl}.\
       \ignorespaces#1
       \vskip\taafter
    \egroup
    \nobreak
    \parindent=0pt
    \let\lasttitle=A%
\everypar={\parindent=\stdparindent
    \penalty\z@\let\lasttitle=N\everypar={}}%
    \ignorespaces}
\def\titleb#1{\par\stepc{Tm}
    \resetcount{Tn}
    \if N\lasttitle\else\vskip\tbbeforeback\fi
    \bgroup
       \normalsize
       \raggedright
       \pretolerance=10000
       \it
       \setbox0=\vbox{\vskip\tbbefore
          \normalsize
          \raggedright
          \pretolerance=10000
          \noindent \it \arabic{Tl}.\arabic{Tm}.\ \ignorespaces#1
          \vskip\tbafter}
       \dimen0=\ht0\advance\dimen0 by\dp0\advance\dimen0 by 2\baselineskip
       \advance\dimen0 by\pagetotal
       \ifdim\dimen0>\pagegoal
          \ifdim\pagetotal>\pagegoal
          \else \if N\lasttitle\eject\fi \fi\fi
       \vskip\tbbefore
       \if N\lasttitle \penalty\subsection@penalty \fi
       \global\subsection@penalty=-100
       \global\subsubsection@penalty=10007
       \noindent \arabic{Tl}.\arabic{Tm}.\ \ignorespaces#1
       \vskip\tbafter
    \egroup
    \nobreak
    \let\lasttitle=B%
    \parindent=0pt
    \everypar={\parindent=\stdparindent
       \penalty\z@\let\lasttitle=N\everypar={}}%
       \ignorespaces}
\def\titlec#1{\par\stepc{Tn}
    \resetcount{To}
    \if N\lasttitle\else\vskip\tcbeforeback\fi
    \bgroup
       \normalsize
       \raggedright
       \pretolerance=10000
       \setbox0=\vbox{\vskip\tcbefore
          \noindent
          \arabic{Tl}.\arabic{Tm}.\arabic{Tn}.\
          \ignorespaces#1\vskip\tcafter}
       \dimen0=\ht0\advance\dimen0 by\dp0\advance\dimen0 by 2\baselineskip
       \advance\dimen0 by\pagetotal
       \ifdim\dimen0>\pagegoal
           \ifdim\pagetotal>\pagegoal
           \else \if N\lasttitle\eject\fi \fi\fi
       \vskip\tcbefore
       \if N\lasttitle \penalty\subsubsection@penalty \fi
       \global\subsubsection@penalty=-50
       \noindent
       \arabic{Tl}.\arabic{Tm}.\arabic{Tn}.\
       \ignorespaces#1\vskip\tcafter
    \egroup
    \nobreak
    \let\lasttitle=C%
    \parindent=0pt
    \everypar={\parindent=\stdparindent
       \penalty\z@\let\lasttitle=N\everypar={}}%
       \ignorespaces}
\def\titled#1{\par\stepc{To}
    \resetcount{Tp}
    \if N\lasttitle\else\vskip\tdbeforeback\fi
    \vskip\tdbefore
    \bgroup
       \normalsize
       \if N\lasttitle \penalty-50 \fi
       \it \noindent \ignorespaces#1\unskip\
    \egroup\ignorespaces}
\def\begref#1{\par
   \unvskip
   \goodbreak\vskip\tabefore
   {\noindent\bf\ignorespaces#1%
   \par\vskip\taafter}\nobreak\let\INS=N}
\def\ref{\if N\INS\let\INS=Y\else\goodbreak\fi
   \hangindent\stdparindent\hangafter=1\noindent\ignorespaces}
\def\endref{\goodbreak}% Ende der Referenzen
\def\acknow#1{\par
   \unvskip
   \vskip\tcbefore
   \noindent{\it Acknowledgements\/}. %
   \ignorespaces#1\par
   \vskip\tcafter}
\def\appendix#1{\vskip\tabefore
    \vbox{\noindent{\bf Appendix #1}\vskip\taafter}%
    \global\eqnum=0\relax
    \nobreak\noindent\ignorespaces}
\let\REFEREE=N
\newbox\refereebox
\setbox\refereebox=\vbox
to0pt{\vskip0.5cm\fullline{\hrulefill\tentt\lower0.5ex
\hbox{\kern5pt referee's copy\kern5pt}\hrulefill}\vss}%
\def\refereelayout{\let\REFEREE=M\footline={\copy\refereebox}
    \message{|A referee's copy will be produced}\par
    \if N\lr\else\if R\lr \onecolumn \fi \let\lr=N \topskip=10pt\fi}

\def\utw{\smash{\rlap{\lower5pt\hbox{$\sim$}}}}
\def\udtw{\smash{\rlap{\lower6pt\hbox{$\approx$}}}}

 %reelle Zahlen
 %natuerliche Zahlen

\def\bbbc{{\mathchoice {\setbox0=\hbox{$\displaystyle\rm C$}\hbox{\hbox
to0pt{\kern0.4\wd0\vrule height0.9\ht0\hss}\box0}}
{\setbox0=\hbox{$\textstyle\rm C$}\hbox{\hbox
to0pt{\kern0.4\wd0\vrule height0.9\ht0\hss}\box0}}
{\setbox0=\hbox{$\scriptstyle\rm C$}\hbox{\hbox
to0pt{\kern0.4\wd0\vrule height0.9\ht0\hss}\box0}}
{\setbox0=\hbox{$\scriptscriptstyle\rm C$}\hbox{\hbox
to0pt{\kern0.4\wd0\vrule height0.9\ht0\hss}\box0}}}}
\def\bbbq{{\mathchoice {\setbox0=\hbox{$\displaystyle\rm Q$}\hbox{\raise
0.15\ht0\hbox to0pt{\kern0.4\wd0\vrule height0.8\ht0\hss}\box0}}
{\setbox0=\hbox{$\textstyle\rm Q$}\hbox{\raise
0.15\ht0\hbox to0pt{\kern0.4\wd0\vrule height0.8\ht0\hss}\box0}}
{\setbox0=\hbox{$\scriptstyle\rm Q$}\hbox{\raise
0.15\ht0\hbox to0pt{\kern0.4\wd0\vrule height0.7\ht0\hss}\box0}}
{\setbox0=\hbox{$\scriptscriptstyle\rm Q$}\hbox{\raise
0.15\ht0\hbox to0pt{\kern0.4\wd0\vrule height0.7\ht0\hss}\box0}}}}
\def\bbbt{{\mathchoice {\setbox0=\hbox{$\displaystyle\rm
T$}\hbox{\hbox to0pt{\kern0.3\wd0\vrule height0.9\ht0\hss}\box0}}
{\setbox0=\hbox{$\textstyle\rm T$}\hbox{\hbox
to0pt{\kern0.3\wd0\vrule height0.9\ht0\hss}\box0}}
{\setbox0=\hbox{$\scriptstyle\rm T$}\hbox{\hbox
to0pt{\kern0.3\wd0\vrule height0.9\ht0\hss}\box0}}
{\setbox0=\hbox{$\scriptscriptstyle\rm T$}\hbox{\hbox
to0pt{\kern0.3\wd0\vrule height0.9\ht0\hss}\box0}}}}
\def\bbbs{{\mathchoice
{\setbox0=\hbox{$\displaystyle     \rm S$}\hbox{\raise0.5\ht0\hbox
to0pt{\kern0.35\wd0\vrule height0.45\ht0\hss}\hbox
to0pt{\kern0.55\wd0\vrule height0.5\ht0\hss}\box0}}
{\setbox0=\hbox{$\textstyle        \rm S$}\hbox{\raise0.5\ht0\hbox
to0pt{\kern0.35\wd0\vrule height0.45\ht0\hss}\hbox
to0pt{\kern0.55\wd0\vrule height0.5\ht0\hss}\box0}}
{\setbox0=\hbox{$\scriptstyle      \rm S$}\hbox{\raise0.5\ht0\hbox
to0pt{\kern0.35\wd0\vrule height0.45\ht0\hss}\raise0.05\ht0\hbox
to0pt{\kern0.5\wd0\vrule height0.45\ht0\hss}\box0}}
{\setbox0=\hbox{$\scriptscriptstyle\rm S$}\hbox{\raise0.5\ht0\hbox
to0pt{\kern0.4\wd0\vrule height0.45\ht0\hss}\raise0.05\ht0\hbox
to0pt{\kern0.55\wd0\vrule height0.45\ht0\hss}\box0}}}}
\def\bbbz{{\mathchoice {\hbox{$\sf\textstyle Z\kern-0.4em Z$}}
{\hbox{$\sf\textstyle Z\kern-0.4em Z$}}
{\hbox{$\sf\scriptstyle Z\kern-0.3em Z$}}
{\hbox{$\sf\scriptscriptstyle Z\kern-0.2em Z$}}}}
\def\diameter{{\ifmmode\oslash\else$\oslash$\fi}}

\def\vec#1{{\boldmath
\textfont0=\bfIXf\scriptfont0=\bfVIf\scriptscriptfont0=\bfVf
\ifmmode
\mathchoice{\hbox{$\displaystyle#1$}}{\hbox{$\textstyle#1$}}
{\hbox{$\scriptstyle#1$}}{\hbox{$\scriptscriptstyle#1$}}\else
$#1$\fi}}
\def\tens#1{\ifmmode
\mathchoice{\hbox{$\displaystyle\sf#1$}}{\hbox{$\textstyle\sf#1$}}
{\hbox{$\scriptstyle\sf#1$}}{\hbox{$\scriptscriptstyle\sf#1$}}\else
$\sf#1$\fi}
\newcount\sterne \sterne=0
\newdimen\fullhead
{\catcode`@=11    % use @ as a normal character
\def\newtoks{\alloc@5\toks\toksdef\@cclvi}
\outer\gdef\makenewtoks#1{\newtoks#1#1={ ????? }}}
\makenewtoks\DATE
\makenewtoks\MAINTITLE
\makenewtoks\SUBTITLE
\makenewtoks\AUTHOR
\makenewtoks\INSTITUTE
\makenewtoks\ABSTRACT
\makenewtoks\KEYWORDS
\makenewtoks\THESAURUS
\makenewtoks\OFFPRINTS
\newlinechar=`\| %
\let\INS=N%
{\catcode`\@=\active
\gdef@#1{\if N\INS $^{#1}$\else\if
E\INS\hangindent0.5\stdparindent\hangafter=1%
\noindent\hbox to0.5\stdparindent{$^{#1}$\hfil}\let\INS=Y\ignorespaces
\else\par\hangindent0.5\stdparindent\hangafter=1
\noindent\hbox to0.5\stdparindent{$^{#1}$\hfil}\ignorespaces\fi\fi}%
}%
\def\mehrsterne{\global\advance\sterne by1\relax}%
\def\footnoterule{\kern-3pt\hrule width 2true cm\kern2.6pt}% Trennlinie
\def\makeOFFPRINTS#1{\bgroup\normalsize
       \hsize=19.5cc
       \baselineskip=10dd\lineskiplimit=0pt\lineskip=0pt
       \def\textindent##1{\noindent{\it Send offprint
          requests to\/}: }\relax
       \vfootnote{nix}{\ignorespaces#1}\egroup}
\def\makesterne{\count254=0\loop\ifnum\count254<\sterne
\advance\count254 by1\star\repeat}
\def\FOOTNOTE#1{\bgroup
       \ifhmode\unskip\fi
       \mehrsterne$^{\makesterne}$\relax
       \normalsize
       \hsize=19.5cc
       \baselineskip=10dd\lineskiplimit=0pt\lineskip=0pt
       \def\textindent##1{\noindent\hbox
       to\stdparindent{##1\hss}}\relax
       \vfootnote{$^{\makesterne}$}{\ignorespaces#1}\egroup}
\def\fonote#1{\ifhmode\unskip\fi
       \mehrsterne$^{\the\sterne}$\bgroup
       \normalsize
       \hsize=19.5cc
       \def\textindent##1{\noindent\hbox
       to\stdparindent{##1\hss}}\relax
       \vfootnote{$^{\the\sterne}$}{\ignorespaces#1}\egroup}
\def\missmsg#1{\message{|Missing #1 }}
\def\tstmiss#1#2#3#4#5{%
\edef\test{\the #1}%
\ifx\test\missing%
  #2\relax%  message
  #3%   action if missing
\else
  \ifx\test\missingi%
    #2\relax%  message
    #3%   action if missing
  \else #4%  action if existing
  \fi
\fi
#5%   action at any rate
}%
\def\maketitle{\paglay%
\def\missing{ ????? }%
\def\missingi{ }%
{\parskip=0pt\relax
\setbox0=\vbox{\hsize=\fullhsize\null\vskip2truecm
\tstmiss%
  {\MAINTITLE}%
  {}%
  {\global\MAINTITLE={MAINTITLE should be given}}%
  {}%
  {%   write MAINTITLE:
   \aTa\ignorespaces\the\MAINTITLE\eTa}%
\tstmiss%
  {\SUBTITLE}%
  {}%
  {}%
  {%   write SUBTITLE:
   \aTb\ignorespaces\the\SUBTITLE\eTb}%
  {}%
\tstmiss%
  {\AUTHOR}%
  {}%
  {\AUTHOR={Name(s) and initial(s) of author(s) should be given}}
  {}%
  {%   write AUTHOR:
\noindent{\bf\ignorespaces\the\AUTHOR\vskip4pt}}%
\tstmiss%
  {\INSTITUTE}%
  {}%
  {\INSTITUTE={Address(es) of author(s) should be given.}}%
  {}%
  {%   write INSTITUTE:
   \let\INS=E
\noindent\ignorespaces\the\INSTITUTE\vskip10pt}%
\tstmiss%
  {\DATE}%
  {}%
  {\DATE={$[$the date of receipt and acceptance should be inserted
later$]$}}%
  {}%
  {%   write DATE:
{\noindent\ignorespaces\the\DATE\vskip21pt}\bf A}%
}%
\global\fullhead=\ht0\global\advance\fullhead by\dp0
\global\advance\fullhead by10pt\global\sterne=0
{\hsize=19.5cc\null\vskip2truecm
\tstmiss%
  {\OFFPRINTS}%
  {}%
  {}%
  {\makeOFFPRINTS{\the\OFFPRINTS}}%
  {}%
\hsize=\fullhsize
\tstmiss%
  {\MAINTITLE}%
  {\missmsg{MAINTITLE}}%
  {\global\MAINTITLE={MAINTITLE should be given}}%
  {}%
  {%   write MAINTITLE:
   \aTa\ignorespaces\the\MAINTITLE\eTa}%
\tstmiss%
  {\SUBTITLE}%
  {}%
  {}%
  {%   write SUBTITLE:
   \aTb\ignorespaces\the\SUBTITLE\eTb}%
  {}%
\tstmiss%
  {\AUTHOR}%
  {\missmsg{name(s) and initial(s) of author(s)}}%
  {\AUTHOR={Name(s) and initial(s) of author(s) should be given}}
  {}%
  {%   write AUTHOR:
\noindent{\bf\ignorespaces\the\AUTHOR\vskip4pt}}%
\tstmiss%
  {\INSTITUTE}%
  {\missmsg{address(es) of author(s)}}%
  {\INSTITUTE={Address(es) of author(s) should be given.}}%
  {}%
  {%   write INSTITUTE:
   \let\INS=E
\noindent\ignorespaces\the\INSTITUTE\vskip10pt}%
\catcode`\@=12
\tstmiss%
  {\DATE}%
  {\message{|The date of receipt and acceptance should be inserted
later.}}%
  {\DATE={$[$the date of receipt and acceptance should be inserted
later$]$}}%
  {}%
  {%   write DATE:
{\noindent\ignorespaces\the\DATE\vskip21pt}}%
}%
\tstmiss%
  {\THESAURUS}%
  {\message{|Thesaurus codes are not given.}}%
  {\global\THESAURUS={missing; you have not inserted them}}%
  {}%
  {}%
\if M\REFEREE\let\REFEREE=Y
\normalbaselineskip=2\normalbaselineskip
\normallineskip=2\normallineskip\normalbaselines\fi
\tstmiss%
  {\ABSTRACT}%
  {\missmsg{ABSTRACT}}%
  {\ABSTRACT={Not yet given.}}%
  {}%
  {\noindent{\bf Abstract. }\ignorespaces\the\ABSTRACT\vskip0.5true cm}%
\def\strich{\par
\vbox to0pt{\hrule width\hsize\vss}\vskip-1.2\baselineskip
\vskip0pt plus3\baselineskip\relax}%
\tstmiss%
  {\KEYWORDS}%
  {\missmsg{KEYWORDS}}%
  {\KEYWORDS={Not yet given.}}%
  {}%
  {\noindent{\bf Key words: }\ignorespaces\the\KEYWORDS
  \strich}%
\global\sterne=0
}}%Ende von maketitle
\newdimen\@txtwd  \@txtwd=\hsize
\newdimen\@txtht  \@txtht=\vsize
\newdimen\@colht  \@colht=\vsize
\newdimen\@colwd  \@colwd=-1pt
\newdimen\@colsavwd
\newcount\in@t \in@t=0
\def\initlr{\if N\lr \ifdim\@colwd<0pt \global\@colwd=\hsize \fi
   \else\global\let\lr=L\ifdim\@colwd<0pt \global\@colwd=\hsize
      \global\divide\@colwd\tw@ \global\advance\@colwd by -10pt
   \fi\fi\global\advance\in@t by 1}
\def\setuplr#1#2#3{\let\lr=O \ifx#1\lr\global\let\lr=N
      \else\global\let\lr=L\fi
   \@txtht=\vsize \@colht=\vsize \@txtwd=#2 \@colwd=#3
   \if N\lr \else\multiply\@colwd\tw@ \fi
   \ifdim\@colwd>\@txtwd\if N\lr
        \errmessage{The text width is less than the column width}%
      \else
        \errmessage{The text width is less the two times the column width}%
      \fi \global\@colwd=\@txtwd
      \if N\lr\divide\@colwd by 2\fi
   \else \global\@colwd=#3 \fi \initlr \@colsavwd=#3
   \global\@insmx=\@txtht
   \global\hsize=\@colwd}
\def\twocolumns{\@fillpage\eject\global\let\lr=L \@makecolht
   \global\@colwd=\@colsavwd \global\hsize=\@colwd}
\def\onecolumn{\@fillpage\eject\global\let\lr=N \@makecolht
   \global\@colwd=\@txtwd \global\hsize=\@colwd}
\def\newpage{\@fillpage\eject}
\def\@fillpage{\vfill\supereject\if R\lr \null\vfill\eject\fi}

\newbox\@leftcolumn
\newbox\@rightcolumn
\newbox\@outputbox
\newbox\@tempboxa
\newbox\@keepboxa
\newbox\@keepboxb
\newbox\@bothcolumns
\newbox\@savetopins
\newbox\@savetopright
\newcount\verybad \verybad=1010
\def\@makecolumn{\ifnum \in@t<1\initlr\fi
   \ifnum\outputpenalty=\the\verybad1  %%% i.e. 10101 if \verybad=1010
      \if L\lr\else\advance\pageno by1\fi
      \message{Warning: There is a 'widow' line
      at the top of page \the\pageno\if R\lr (left)\fi.
      This is unacceptable.} \if L\lr\else\advance\pageno by-1\fi \fi
   \ifnum\outputpenalty=\the\verybad2
      \message{Warning: There is a 'club' line
      at the bottom of page \the\pageno\if L\lr(left)\fi.
      This is unacceptable.} \fi
   \if L\lr \ifvoid\@savetopins\else\@colht=\@txtht\fi \fi
   \if R\lr \ifvoid\@bothcolumns \ifvoid\@savetopright
       \else\@colht=\@txtht\fi\fi\fi
   \global\setbox\@outputbox
   \vbox to\@colht{\boxmaxdepth\maxdepth
   \if L\lr \ifvoid\@savetopins\else\unvbox\@savetopins\fi \fi
   \if R\lr \ifvoid\@bothcolumns \ifvoid\@savetopright\else
       \unvbox\@savetopright\fi\fi\fi
   \ifvoid\topins\else\ifnum\count\topins>0
         \ifdim\ht\topins>\@colht
            \message{|Error: Too many or too large single column
            box(es) on this page.}\fi
         \unvbox\topins
      \else
         \global\setbox\@savetopins=\vbox{\ifvoid\@savetopins\else
         \unvbox\@savetopins\penalty-500\fi \unvbox\topins} \fi\fi
   \dimen@=\dp\@cclv \unvbox\@cclv % open up \box255
   \ifvoid\bottomins\else\unvbox\bottomins\fi
   \ifvoid\footins\else % footnote info is present
     \vskip\skip\footins
     \footnoterule
     \unvbox\footins\fi
   \ifr@ggedbottom \kern-\dimen@ \vfil \fi}%
}
\def\@outputpage{\@dooutput{\lr}}
\def\@colbox#1{\hbox to\@colwd{\box#1\hss}}
\def\@dooutput#1{\global\topskip=10pt
  \ifdim\ht\@bothcolumns>\@txtht
    \if #1N
       \unvbox\@outputbox
    \else
       \unvbox\@leftcolumn\unvbox\@outputbox
    \fi
    \global\setbox\@tempboxa\vbox{\hsize=\@txtwd\makeheadline
       \vsplit\@bothcolumns to\@txtht
       \makefootline\hsize=\@colwd}%
    \message{|Error: Too many double column boxes on this page.}%
    \shipout\box\@tempboxa\advancepageno
    \unvbox255 \penalty\outputpenalty
  \else
    \global\setbox\@tempboxa\vbox{\hsize=\@txtwd\makeheadline
       \ifvoid\@bothcolumns\else\unvbox\@bothcolumns\fi
       \hsize=\@colwd
       \if #1N
          \hbox to\@txtwd{\@colbox{\@outputbox}\hfil}%
       \else
          \hbox to\@txtwd{\@colbox{\@leftcolumn}\hfil\@colbox{\@outputbox}}%
       \fi
       \hsize=\@txtwd\makefootline\hsize=\@colwd}%
    \shipout\box\@tempboxa\advancepageno
  \fi
  \ifnum \special@pages>0 \s@count=100 \page@command
      \xdef\page@command{}\global\special@pages=0 \fi
  }
\def\balance@right@left{\dimen@=\ht\@leftcolumn
    \advance\dimen@ by\ht\@outputbox
    \advance\dimen@ by\ht\springer@macro
    \dimen2=\z@ \global\the@end=0
    \ifdim\dimen@>70pt\setbox\z@=\vbox{\unvbox\@leftcolumn
          \unvbox\@outputbox}%
       \loop
          \dimen@=\ht\z@
          \advance\dimen@ by0.5\topskip
          \advance\dimen@ by\baselineskip
          \advance\dimen@ by\ht\springer@macro
          \advance\dimen@ by\dimen2
          \divide\dimen@ by2
          \splittopskip=\topskip
          % Now split it to two parts of about the same height
          {\vbadness=10000
             \global\setbox3=\copy\z@
             \global\setbox1=\vsplit3 to \dimen@}%
          \dimen1=\ht3 \advance\dimen1 by\ht\springer@macro
       \ifdim\dimen1>\ht1 \advance\dimen2 by\baselineskip\repeat
       \dimen@=\ht1
       % Restore the column boxes and adjust
       \global\setbox\@leftcolumn
          \hbox to\@colwd{\vbox to\@colht{\vbox to\dimen@{\unvbox1}\vfil}}%
       \global\setbox\@outputbox
          \hbox to\@colwd{\vbox to\@colht{\vbox to\dimen@{\unvbox3
             \vfill\box\springer@macro}\vfil}}%
    \else
       \setbox\@leftcolumn=\vbox{unvbox\@leftcolumn\bigskip
          \box\springer@macro}%
    \fi}
\newinsert\bothins
\newbox\rightins
\skip\bothins=\z@skip
\count\bothins=1000
\dimen\bothins=\@txtht \advance\dimen\bothins by -\bigskipamount
\def\bothtopinsert{\par\begingroup\setbox\z@\vbox\bgroup
    \hsize=\@txtwd\parskip=0pt\par\noindent\bgroup}
\def\endbothinsert{\egroup\egroup
  \if R\lr
    \right@nsert
  \else    % L\lr or N\lr
    \dimen@=\ht\z@ \advance\dimen@ by\dp\z@ \advance\dimen@ by\pagetotal
    \advance\dimen@ by \bigskipamount \advance\dimen@ by \topskip
    \advance\dimen@ by\ht\topins \advance\dimen@ by\dp\topins
    \advance\dimen@ by\ht\bottomins \advance\dimen@ by\dp\bottomins
    \advance\dimen@ by\ht\@savetopins \advance\dimen@ by\dp\@savetopins
    \ifdim\dimen@>\@colht\right@nsert\else\left@nsert\fi
  \fi  \endgroup}
\def\right@nsert{\global\setbox\rightins\vbox{\ifvoid\rightins
    \else\unvbox\rightins\fi\penalty100
    \splittopskip=\topskip
    \splitmaxdepth\maxdimen \floatingpenalty200
    \dimen@\ht\z@ \advance\dimen@\dp\z@
    \box\z@\nobreak\bigskip}}
\def\left@nsert{\insert\bothins{\penalty100
    \splittopskip=\topskip
    \splitmaxdepth\maxdimen \floatingpenalty200
    \box\z@\nobreak\bigskip}
    \@makecolht}
\newdimen\@insht    \@insht=\z@
\newdimen\@insmx    \@insmx=\vsize
\def\@makecolht{\global\@colht=\@txtht \@compinsht
    \global\advance\@colht by -\@insht \global\vsize=\@colht
    \global\dimen\topins=\@colht}
\def\@compinsht{\if R\lr
       \dimen@=\ht\@bothcolumns \advance\dimen@ by\dp\@bothcolumns
       \ifvoid\@bothcolumns \advance\dimen@ by\ht\@savetopright
          \advance\dimen@ by\dp\@savetopright \fi
    \else
       \dimen@=\ht\bothins \advance\dimen@ by\dp\bothins
       \advance\dimen@ by\ht\@savetopins \advance\dimen@ by\dp\@savetopins
    \fi
    \ifdim\dimen@>\@insmx
       \global\@insht=\dimen@
    \else\global\@insht=\dimen@
    \fi}
\newinsert\bottomins
\skip\bottomins=\z@skip
\count\bottomins=1000
\xdef\page@command{}
\newcount\s@count
\newcount\special@pages \special@pages=0
\def\specialpage#1{\global\advance\special@pages by1
    \global\s@count=\special@pages
    \global\advance\s@count by 100
    \global\setbox\s@count
    \vbox to\@txtht{\hsize=\@txtwd\parskip=0pt
    \par\noindent\noexpand#1\vfil}%
    \def\protect{\noexpand\protect\noexpand}%
    \xdef\page@command{\page@command
         \protect\global\advance\s@count by1
         \protect\begingroup
         \protect\setbox\z@\vbox{\protect\makeheadline
                                    \protect\box\s@count
            \protect\makefootline}%
         \protect{\shipout\box\z@}%
         \protect\endgroup\protect\advancepageno}%
    \let\protect=\relax
   }
\def\@startins{\vskip \topskip\hrule height\z@
   \nobreak\vskip -\topskip\vskip3.7pt}
\let\retry=N
\output={\@makecolht \global\topskip=10pt \let\retry=N%
   \ifnum\count\topins>0 \ifdim\ht\topins>\@colht
       \global\count\topins=0 \global\let\retry=Y%
       \unvbox\@cclv \penalty\outputpenalty \fi\fi
   \if N\retry
    \if N\lr     % this is for single column output
       \@makecolumn
       \ifnum\the@end>0
          \setbox\z@=\vbox{\unvcopy\@outputbox}%
          \dimen@=\ht\z@ \advance\dimen@ by\ht\springer@macro
          \ifdim\dimen@<\@colht
             \setbox\@outputbox=\vbox to\@colht{\box\z@
             \unskip\vskip12pt plus0pt minus12pt
             \box\springer@macro\vfil}%
          \else \box\springer@macro \fi
          \global\the@end=0
       \fi
       \ifvoid\bothins\else\global\setbox\@bothcolumns\box\bothins\fi
       \@outputpage
       \ifvoid\rightins\else
       %  Hold \rightins back if there is already a \@savetopins
       \ifvoid\@savetopins\insert\bothins{\unvbox\rightins}\fi
       \fi
    \else
       \if L\lr    % this is the left of two columns
          \@makecolumn
          \global\setbox\@leftcolumn\box\@outputbox \global\let\lr=R%
          \ifnum\pageno=1
             \message{|[left\the\pageno]}%
          \else
             \message{[left\the\pageno]}\fi
          \ifvoid\bothins\else\global\setbox\@bothcolumns\box\bothins\fi
          \global\dimen\bothins=\z@
          \global\count\bothins=0
          \ifnum\pageno=1
             \global\topskip=\fullhead\fi
       \else    % the right column
          \@makecolumn
          \ifnum\the@end>0\ifnum\pageno>1\balance@right@left\fi\fi
          \@outputpage \global\let\lr=L%
          \global\dimen\bothins=\maxdimen
          \global\count\bothins=1000
          \ifvoid\rightins\else
          %  Hold \rightins back if there is already a \@savetopins
             \ifvoid\@savetopins \insert\bothins{\unvbox\rightins}\fi
          \fi
       \fi
    \fi
    \global\let\last@insert=N \put@default
    \ifnum\outputpenalty>-\@MM\else\dosupereject\fi
    \ifvoid\@savetopins\else
      \ifdim\ht\@savetopins>\@txtht
        \global\setbox\@tempboxa=\box\@savetopins
        \global\setbox\@savetopins=\vsplit\@tempboxa to\@txtht
        \global\setbox\@savetopins=\vbox{\unvbox\@savetopins}%
        \global\setbox\@savetopright=\box\@tempboxa \fi
    \fi
    \@makecolht
    \global\count\topins=1000
   \fi
   }
\if N\lr
   \setuplr{O}{\fullhsize}{\hsize}% O = one column
\else
   \setuplr{T}{\fullhsize}{\hsize}% T = two columns
\fi
\def\put@default{\global\let\insert@here=Y
   \global\let\insert@at@the@bottom=N}%
\def\puthere{\global\let\insert@here=Y%
    \global\let\insert@at@the@bottom=N}
\def\putattop{\global\let\insert@here=N%
    \global\let\insert@at@the@bottom=N}
\def\putatbottom{\global\let\insert@here=N%
    \global\let\insert@at@the@bottom=X}
\put@default
\let\last@insert=N
\def\end@skip{\smallskip}
\newdimen\min@top
\newdimen\min@here
\newdimen\min@bot
\min@top=10cm
\min@here=4cm
\min@bot=\topskip
\def\figfuzz{\vskip 0pt plus 6pt minus 3pt}  % more flexible spacing
%--------------------------------------------------------------------
\def\check@here@and@bottom#1{\relax
   \ifvoid\topins\else       \global\let\insert@here=N\fi
   \if B\last@insert         \global\let\insert@here=N\fi
   \if T\last@insert         \global\let\insert@here=N\fi
   \ifdim #1<\min@bot        \global\let\insert@here=N\fi
   \ifdim\pagetotal>\@colht  \global\let\insert@here=N\fi
   \ifdim\pagetotal<\min@here\global\let\insert@here=N\fi
   \if X\insert@at@the@bottom\global\let\insert@at@the@bottom=Y
     \else\if T\last@insert  \global\let\insert@at@the@bottom=N\fi
          \if H\last@insert  \global\let\insert@at@the@bottom=N\fi
          \ifvoid\topins\else\global\let\insert@at@the@bottom=N\fi\fi
   \ifdim #1<\min@bot        \global\let\insert@at@the@bottom=N\fi
   \ifdim\pagetotal>\@colht  \global\let\insert@at@the@bottom=N\fi
   \ifdim\pagetotal<\min@top \global\let\insert@at@the@bottom=N\fi
   \ifvoid\bottomins\else    \global\let\insert@at@the@bottom=Y\fi
   \if Y\insert@at@the@bottom\global\let\insert@here=N\fi }
\def\single@column@insert#1{\relax
   \setbox\@tempboxa=\vbox{#1}%
   \dimen@=\@colht \advance\dimen@ by -\pagetotal
   \advance\dimen@ by-\ht\@tempboxa \advance\dimen0 by-\dp\@tempboxa
   \advance\dimen@ by-\ht\topins \advance\dimen0 by-\dp\topins
   \check@here@and@bottom{\dimen@}%
   \if Y\insert@here
      \par  % The insertion forces a new paragraph in this case.
      \midinsert\figfuzz\relax     %%%%%%%%%\bigskip
      \box\@tempboxa\end@skip\figfuzz\endinsert
      \global\let\last@insert=H
   \else \if Y\insert@at@the@bottom
      \begingroup\insert\bottomins\bgroup\if B\last@insert\end@skip\fi
      \floatingpenalty=20000\figfuzz\bigskip\box\@tempboxa\egroup\endgroup
      \global\let\last@insert=B
   \else
      \topinsert\box\@tempboxa\end@skip\figfuzz\endinsert
      \global\let\last@insert=T
   \fi\fi\put@default\ignorespaces}
\def\begfig#1cm#2\endfig{\single@column@insert{\@startins\rahmen{#1}#2}%
\ignorespaces}
\def\begfigwid#1cm#2\endfig{\relax
   \if N\lr  % Here the only difference to \begfig is the larger \hsize
      {\hsize=\fullhsize \begfig#1cm#2\endfig}%
   \else
      \setbox0=\vbox{\hsize=\fullhsize\bigskip#2\smallskip}%
      \dimen0=\ht0\advance\dimen0 by\dp0
      \advance\dimen0 by#1cm
      \advance\dimen0by7\normalbaselineskip\relax
      \ifdim\dimen0>\@txtht
         \message{|Figure plus legend too high, will try to put it on a
                  separate page. }%
         \begfigpage#1cm#2\endfig
      \else
         \bothtopinsert\line{\vbox{\hsize=\fullhsize
         \@startins\rahmen{#1}#2\smallskip}\hss}\figfuzz\endbothinsert
      \fi
   \fi}
\def\begfigside#1cm#2cm#3\endfig{\relax
   \if N\lr  % Here the only difference to \begfig is the larger \hsize
      {\hsize=\fullhsize \begfig#1cm#3\endfig}%
   \else
      \dimen0=#2true cm\relax
      \ifdim\dimen0<\hsize
         \message{|Your figure fits in a single column; why don't|you use
                  \string\begfig\space instead of \string\begfigside? }%
      \fi
      \dimen0=\fullhsize
      \advance\dimen0 by-#2true cm
      \advance\dimen0 by-1true cc\relax
      \bgroup
         \ifdim\dimen0<8true cc\relax
            \message{|No sufficient room for the legend;
                     using \string\begfigwid. }%
            \begfigwid #1cm#3\endfig
         \else
            \ifdim\dimen0<10true cc\relax
               \message{|Room for legend to narrow;
                        legend will be set raggedright. }%
               \rightskip=0pt plus 2cm\relax
            \fi
            \setbox0=\vbox{\def\figure##1##2{\vbox{\hsize=\dimen0\relax
                           \@startins\noindent\petit{\bf
                           Fig.\ts##1\unskip.\ }\ignorespaces##2\par}}%
                           #3\unskip}%
            \ifdim#1true cm<\ht0\relax
               \message{|Text of legend higher than figure; using
                        \string\begfig. }%
               \begfigwid #1cm#3\endfig
            \else
               \def\figure##1##2{\vbox{\hsize=\dimen0\relax
                                       \@startins\noindent\petit{\bf
                                       Fig.\ts##1\unskip.\
                                       }\ignorespaces##2\par}}%
               \bothtopinsert\line{\vbox{\hsize=#2true cm\relax
               \@startins\rahmen{#1}}\hss#3\unskip}\figfuzz\endbothinsert
            \fi
         \fi
      \egroup
   \fi\ignorespaces}
\def\begfigpage#1cm#2\endfig{\specialpage{\@startins
   \vskip3.7pt\rahmen{#1}#2}\ignorespaces}%
\def\begtab#1cm#2\endtab{\single@column@insert{#2\rahmen{#1}}\ignorespaces}
\let\begtabempty=\begtab
\def\begtabfull#1\endtab{\single@column@insert{#1}\ignorespaces}
\def\begtabemptywid#1cm#2\endtab{\relax
   \if N\lr
      {\hsize=\fullhsize \begtabempty#1cm#2\endtab}%
   \else
      \bothtopinsert\line{\vbox{\hsize=\fullhsize
      #2\rahmen{#1}}\hss}\medskip\endbothinsert
   \fi\ignorespaces}
\def\begtabfullwid#1\endtab{\relax
   \if N\lr
      {\hsize=\fullhsize \begtabfull#1\endtab}%
   \else
      \bothtopinsert\line{\vbox{\hsize=\fullhsize
      \noindent#1}\hss}\medskip\endbothinsert
   \fi\ignorespaces}
\def\begtabpage#1\endtab{\specialpage{#1}\ignorespaces}
\catcode`\@=\active   % This is reset by the \maketitle macro
%
%\def\spose#1{\hbox to 0pt{#1\hss}}
%\def\Dt{\spose{\raise 1.5ex\hbox{\hskip3pt$\mathchar"201$}}} % upper case
%\def\dt{\spose{\raise 1.0ex\hbox{\hskip2pt$\mathchar"201$}}} % lower case
%
%\input $1$dia0:[soft.aatex]aa.cmm
%\refereelayout
%
\MAINTITLE={
A ROSAT HRI observation of the cooling flow cluster MS0839.9+2938}
%\FOOTNOTE{ based on observations of the ROSAT satellite} }

\AUTHOR={
 R. Nesci@1, G.C. Perola@{1,2} and A. Wolter@3
}
\OFFPRINTS={ R. Nesci  }

\INSTITUTE={
@1 Istituto Astronomico, Universita' di Roma La Sapienza, Via G. M. Lancisi
29, I-00161 Roma

@2 Dipartimento di Fisica, III Universita' di Roma

@3 Osservatorio Astronomico di Brera, Via Brera 28, I-20121 Milano
}
\DATE{ Received ....; accepted ....}

\ABSTRACT{A ROSAT HRI observation of the cluster MS\-0839.9+2938
at z=0.194 is presented. It confirms the earlier suggestion,
based on the detection of extended H$\alpha$ emission, that the inner regions
of this cluster are dominated by a cooling flow. The surface brightness
distribution within the cooling radius shows structures which can
be interpreted as evidence of a highly inhomogeneous space distribution
of the cooling process. We note that the brightness at the barycentre
of its distribution, which falls right on top of the central giant
elliptical galaxy, is lower than the peaks in the structures around:
we suggest that this situation is likely to arise as a consequence
of photoelectric absorption by cold gas within the cooling flow, with
an equivalent column density in the order of 5$\times$10$^{21}$ cm$^{-2}$
within $\sim$10" from the centre.
}
\KEYWORDS{ Galaxies: clustering; cooling flows; Galaxies: clusters: individual:
 MS0839+2938 }
\THESAURUS{ 11.03.01; 11.03.3; 11.03.04 }
\maketitle

\titlea {Introduction}

The cluster MS0839+2938 was discovered
as an extended X-ray source in the EINSTEIN {\it Extended
Medium Sensitivity Survey}
(Gioia et al. 1987, 1990). Subsequent optical and radio observations
(Nesci et al. 1989) showed that the cluster is
domina\-ted by a giant elliptical, probably a cD type galaxy,
which is characterized by an intense and extended
H$\alpha$ emission, and a compact, relatively weak radio emission.
The redshift of this galaxy is z=0.193, and the cluster galaxy velocity
dispersion along the line of sight, estimated on the basis of
six redshifts, is 1500$\pm$700 km s$^{-1}$.
At the distance of 1200 Mpc (H$_0$=50 km s$^{-1}$ Mpc$^{-1}$, q$_0$ = 0.5),
the X-ray luminosity in the 0.3-3.5 keV band was estimated to be about
4$\times10^{44}$ erg s$^{-1}$.

Since a spatially extended optical line emission is present in several
nearby cooling flow
clusters, MS0839+2938 appeared to be one of the most distant candidate
cooling flow cluster known at the epoch of the ROSAT launch.
Strictly speaking, the identification of a cooling flow
would require the detection in the X-ray emitting gas
of a much lower temperature
in the innermost regions  than in the whole cluster.
However, when this goal cannot be directly achieved,
a cooling flow regime is generally regarded as proven if the central
density, as derived from the cluster surface brightness distribution
adopting the dominant cluster temperature, implies a cooling time
definitely shorter than the Hubble time (see e.g. Arnaud 1988).
The aim of our ROSAT observation was therefore
to derive the density profile of MS0839+2938 and verify the
existence of the cooling flow. Since the scale at the cluster distance
is 4.13 kpc arcsec$^{-1}$, in order to resolve the emission within the core
radius we used the HRI instrument. This choice, however, implied the
loss of temperature information.
A detailed description of the satellite and its instruments can be found
in the ``ROSAT AO2 Call for Proposals" (MPE, 1991)

\titlea{ Data analysis}

The pointed observation of the cluster MS0839 +2938 (image number WG800159H.N2)
lasted 18879 live seconds and was performed on April 20-22, 1992. Besides the
extended image of the cluster at the center of the field of view, a point
source was also detected 4.7' to the SW (32 net cts), corresponding to the
star HD74076;
this allowed us to check the zero point of the sky coordinates of our image
and to overlay the X-ray onto the optical image of the cluster
with the accuracy of $\sim$2".

We determined the background level by
measuring the counts in 32 adjacent squares with sides of 260" all
around the cluster in the central, 1600" side
square of the frame. The average count rate
is 1.150$\times$10$^{-6}$ cts arcsec$^{-2}$ s$^{-1}$
(0.0217 cts arcsec$^{-2}$), very near to the  pre-flight
expected value of 1.1$\times 10^{-6}$ and to the value ($1.25 \times 10^{-6}$)
found by
Sarazin et al. (1992a) in their ROSAT HRI observation of 2A 0335+098.
The standard deviation of the 32 measurements is
fully consistent with the Poisson statistics. We
also measured the background level in
a ring 200"-280" around the cluster center finding the same result.

\titlea{ X-ray morphology of the cluster}

In order to evaluate the maximum extent out to which the cluster emission
is detected, after background subtraction an isophotal map was obtained
by smoothing the original data, binned in 1"$\times$1" pixels, through a
Gaussian filter with $\sigma$ = 32". The map is shown in Fig. 1, where the
lowest contour corresponds to four times the background fluctuations:
at this brightness level (0.005 cts arcsec$^{-2}$) the emission
is elongated in the N-S direction, as the dominant
galaxy, with a major axis of $\sim$5 arcmin ($\sim$1200 kpc).
An approximately circular type of symmetry is evident in the less
smoothed Fig. 2,
obtained through a Gaussian filter with $\sigma$ = 8", which shows
the emission to be centrally peaked, with the peak coinciding
with the dominant galaxy, thus leading to a morphological classification
as an XD cluster, according to the scheme proposed by Jones and Forman (1984).
%{\bf  e qui c'e' la figura uno che e' di una colonna e alta 8cm}
%
%  figura a una sola colonna alta 8 cm
%

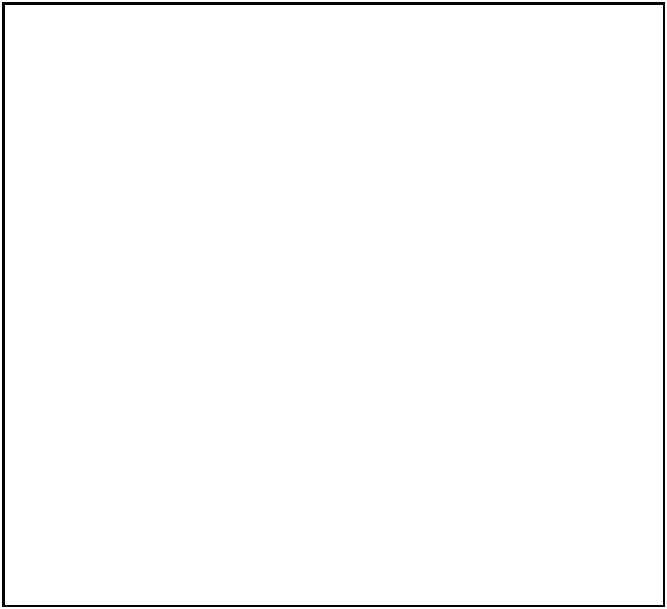
\begfig 8cm \figure{1}{The X-ray map at low resolution after background
subtraction. The isophotes are at levels of 0.0051, 0.0076, 0.0102, 0.0127,
0.0191, 0.0254, 0.0317, 0.0381, 0.0508 net cts arcsec$^{-2}$.}
\endfig

% The isophotes are logarithmically spaced at steps of 0.20: the lowest value
%%is
% 0.006 net counts arcsec$^{-2}$. The heavy lines include regions below the
% assumed average background value. }

%{\bf  e qui c'e' la figura due che e' di una colonna e alta 8cm}
%
%  figura a una sola colonna alta 8 cm
%
   \begfig 8cm \figure{2}
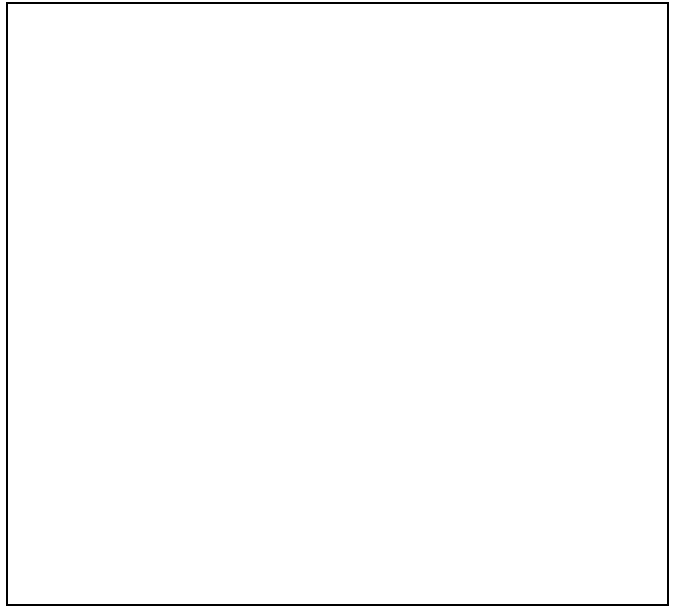
{The isophotal map at intermediate resolution:
the isophotes are at levels of 0.024, 0.044, 0.064, 0.084, 0.104, 0.144, 0.184
and 0.264 net cts arcsec$^{-2}$. }
\endfig

The isophotal map at high resolution (unbinned and smoo\-thed with a Gaussian
filter with $\sigma$
=2") of the very cen\-tral region is shown in Fig.3 overlaid onto an optical
(R)
image of the cluster (Nesci et al. 1989).
This map shows that the brightness distribution in the central region is
not smooth, but that it contains structures with sizes in the range
of a few tens
of kpc. The brightness contrast and the length scale of these structures
are consistent, taking into account the blurring due the difference in
distance, with those discovered with the ROSAT HRI in two relatively nearby
cooling flow clusters, 2A0335+096 (z=0.035) and A2029 (z=0.0767) by
Sarazin et al. (1992a, b). Given the limited count statistics, the details
in the observed structures, such as their exact brightness excess, can be
determined only with marginal significance, except for the bright and
resolved blob located about 12" to the South of the giant elliptical:
the peak excess brightness of this blob is approximately equal to that of the
surrounding diffuse emission, and the size is about 6", or 24 kpc, across.

Another important morphological aspect is that the brightness,
at the position of the giant elliptical galaxy,
is lower than the peaks in the surrounding structures,
while the radio position of the galaxy centre (8$^h$42$^m$55.9$^s$;
+29$^\circ$27'27" (Eq. 2000)) coincides within about 2"
with the barycentre of the overall X-ray
emission. We shall return on these aspects in Sections 4 and 5.

As a side issue, we note that no X-rays were detected from the emission
line galaxy (G27 in Nesci et al. 1989) with a very blue colour (V-R=0.42),
located 68" West of the dominant cluster galaxy and at a similar redshift
(z=0.186).

%{\bf  e qui c'e' la figura tre che e' di una colonna e alta 8cm}
%
%  figura a una sola colonna alta 8 cm
%
   \begfig 8cm \figure{3}
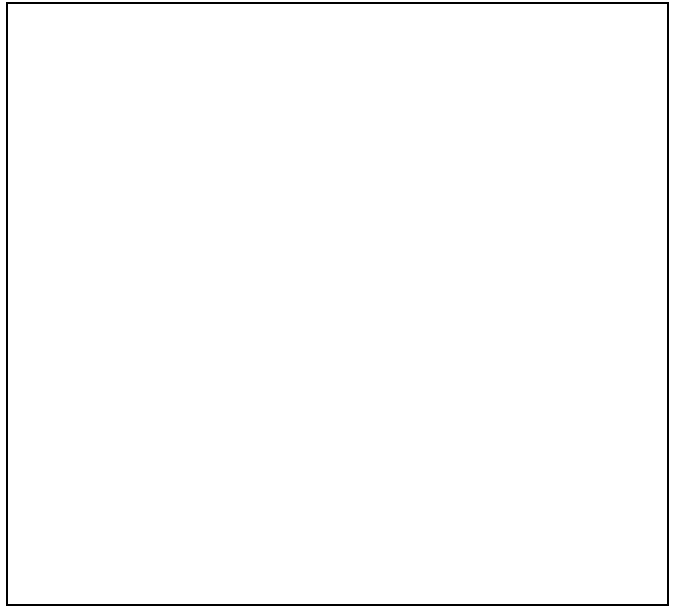
{Map at the highest resolution of the inner cluster region overlaid onto
to the optical
image of the cluster. The isophotal contours are 0.1383, 0.2183, 0.2983,
0.3783,
0.4583, 0.5383 and 0.6183 net cts arcsec$^{-2}$.}
\endfig

\titlea{ Physical parameters}

The radial profile of the brightness distribution, azimuthally averaged in
steps of 10", is given in Fig. 4. The profile beyond 50" ($\sim$200 kpc) can be
well fitted with a power law of slope s=$-2.12\pm$0.42 at the 95$\%$
confidence level. This slope corresponds to a value for the
$\beta$  parameter of the hydrostatic isothermal model (Cavaliere
and Fusco-Femiano 1976) of 0.52$\pm$0.07, similar to
those found by Sarazin et al. (1992a,b) using the ROSAT HRI for 2A 0335+096
(0.55) and A2029 (0.61).

%{\bf  e qui c'e' la figura quattro che e' di una colonna e alta 8cm}
%
%  figura a una sola colonna alta 8 cm
%
   \begfig 8cm \figure{4}
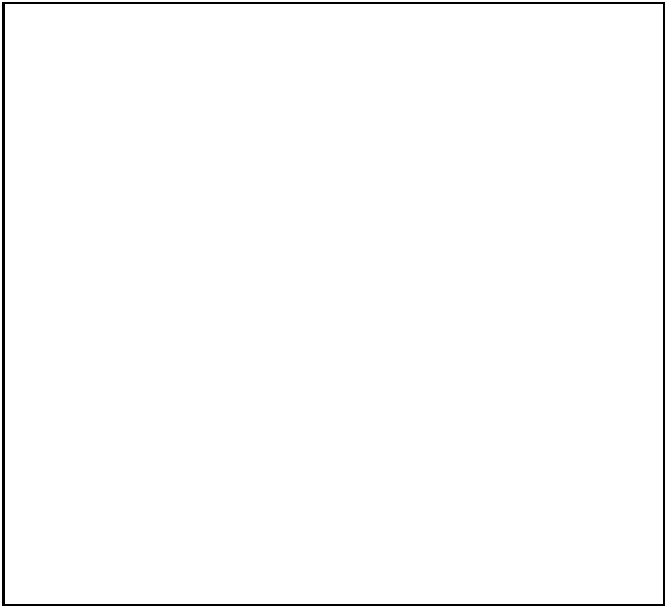
{The profile of the surface brightness {\it S} azimuthally averaged in rings
of 10", as a function of radius {\it r}.
The best fitting power-law profile beyond of 50" is shown as a
solid line. Error bars represent the 68$\%$ confidence level in Poisson
statistics.
}
\endfig

The conversion from observed counts to physical flux requires an
assumption for the X-ray emission spectrum of the cluster and an
estimate of the absorption along the line of sight. For the plasma
thermal spectrum we adopted the code by Mewe et al. (1985)
with 0.3 solar abundance, a typical value for intracluster gas (Edge et al.
1991a), and
folded it with the effective area of the ROSAT HRI.
A correction was also applied to take into account the energy band shift
due to the cluster redshift (k-correction), but this correction is very small
with respect to
that due to the galactic absorption. For the latter a hydrogen column
density of log(N$_H$)=20.6  was adopted, as in Nesci et al. (1989).

The gas temperature T can be evaluated using the empirical correlation
(Edge and Stewart 1991b)
between T and the radial velocity dispersion of the cluster galaxies.
However, given the
small number (6) of redshifts available, the dispersion estimate
is subject to a large uncertainty related to projection, and possibly even
contamination effects. We therefore preferred to adopt two
rather different values of kT (3 and 7 keV) and to check
`a posteriori' whether the derived bolometric
cluster luminosity fits in the empirical luminosity-temperature relation
(Edge and Stewart 1991a).

The absorption corrected flux within 120" from the cluster centre,
in the band 0.1-2.4 keV,
is (2.45$\pm$0.11)$\times$10$^{-12}$ erg cm$^{-2}$ s$^{-1}$ for kT=7 keV,
and (2.50$\pm$0.11)$\times$10$^{-12}$ for kT=3 keV: these values are
consistent with the estimate based on the EINSTEIN
Observatory observation within the same radius, (2.23$\pm$0.24) $\times$
10$^{-12}$ erg cm$^{-2}$ s$^{-1}$ rescaled to the ROSAT band.
Within a radius of 200", which encircles the lowest contour of Fig. 1,
and contains 948 net counts,
the total flux is (2.83$\pm$0.14) $\times10^{-12}$ erg cm$^{-2}$ s$^{-1}$
for kT=7 keV, and 2.86$\times$10$^{-12}$ for kT=3 keV.
This corresponds to a luminosity in the ROSAT band of 4.8$\times$10$^{44}$
erg s$^{-1}$ for the temperature range considered; the bolometric luminosity
then is 1.25$\times$10$^{45}$ erg s$^{-1}$ if kT=7 keV, or
7.5$\times$10$^{44}$ erg s$^{-1}$ if kT=3 keV.

Edge and Stewart (1991a, their Fig. 11), showed that there exists a
correlation between the temperature and the bolome\-tric luminosity, whose
normalization however depends on the value of the central density.
As we shall presently see, the central density derived from the X-ray data
is rather insensitive to
which one of the two temperature values is adopted, and our cluster
would belong to their subsample of objects with central density
$>$ 9$\times10^{-3}$ cm$^{-3}$. With respect to their best fitting correlation
line in the kT-L$_{bol}$ plane, this cluster falls to the right (its
luminosity is too low) if kT=7 keV, to the left (its luminosity is too
high) if kT=3 keV. The actual cluster temperature is therefore
likely to be in this range. In the following we shall derive
the relevant parameters for both cases.

The cluster one dimensional velocity dispersion expected from the derived value
of $\beta$ is $\sim$800 km s$^{-1}$ for kT=7 keV and $\sim$520 km s$^{-1}$ for
kT=3 keV. The estimate based on six redshifts and reported in Section 1
is marginally consistent with the first value, but it could be
overestimated due to the contamination by a
non-member object. After removing the most discrepant redshift we obtain
a dispersion of 850 ($\pm$400) km s$^{-1}$, quite consistent with either of
the two values derived above.

{}From the observed surface brightness profile,
assuming spherical symmetry and a constant temperature,
it is possible to derive the density profile with a deprojection
technique (see e.g. Arnaud 1988). To
overcome numerical problems arising from the poor statistics on the points
at large radii, we deprojected an analytical fit
of the brightness profile. The most relevant resulting parameters,
namely the central density n$_0$, the central cooling time t$_{co}$, the
cooling radius r$_c$ (defined as the radius where the cooling time equals
the Hubble time minus the look back time,
that is $10^{10}$ years), are given in Table 1 for
the two temperatures; the density profile is illustrated in Fig. 5.
\medskip
\hbox{Table 1: Physical parameters}
{\hrule}
{\medskip}
\hbox{~~kT~~~~~~~n$_0$~~~~~~~~~t$_{co}$~~~~~~~~r$_c$~~~~~~~M$_{gas}$($\leq$0.5
Mpc)}
\hbox{~(keV)~~~(cm$^{-3}$)~~(Gyrs)~~~(kpc)~~~~~(M$_\odot$)}
{\medskip}
{\hrule}
{\medskip}
\hbox{~~~7~~~~~0.0170~~~~~~3.96~~~~~~~70~~~~~~1.1$\times$10$^{13}$}
\hbox{~~~3~~~~~0.0156~~~~~~2.57~~~~~~110~~~~~~1.2$\times$10$^{13}$}
{\medskip}
{\hrule}
\medskip
{}From this table it can be seen that the derived central density depends very
weakly on the adopted temperature. Furthermore, no matter which one of the two
temperature values is adopted, the central cooling time of MS0839+2938
is definitely lower than the Hubble time, so this cluster certainly
contains a cooling flow.
The mass flow rate within the cooling radius can be estimated from the
relation
$\Dt M=0.4 \mu m_H L_{bol}/(kT)$, where $\mu$ is the mean molecular gas weight,
L$_{bol}$ is the bolome\-tric X-ray luminosity within the cooling
radius and m$_H$ is the hydrogen mass.
We derive $\Dt M$=130 M$_\odot$ y$^{-1}$ for kT=7
keV and $\Dt M$=280 M$_\odot$ y$^{-1}$ for kT=3 keV.

It is noteworthy that the map in Fig. 3 is dominated by emission from the
cooling flow region, hence the structures vi\-si\-ble in the map lie within the
cooling radius and are likely to be the result of inhomogeneities where the
cooling process goes on more rapidly. By assuming that these structures are
in pressure equilibrium with the surrounding medium, one can estimate their
temperature and therefore their proper cooling time. Let us for instance
consider the bright blob to the South of the central galaxy, and assume that it
has a spherical configuration
with radius 12 kpc and a filling factor equal one. By adopting a
distance from the centre equal to the projected distance, from Fig. 5 one
obtains the pressure term nT; for an external temperature of 7 keV, it
turns out that the observed excess counts in the blob require that the
internal temperature be about 3.5 keV, and the internal density therefore
be twice the external value. The cooling time in the blob is then about
2$\times$10$^9$ years, a factor of four shorter than that of the surrounding
gas. The temperature would be lower, and the cooling time shorter, if the
filling factor were less than unity.

%{\bf  e qui c'e' la figura cinque che e' di una colonna e alta 8cm}
%
%  figura a una sola colonna alta 8 cm
%
   \begfig 8cm \figure{5}
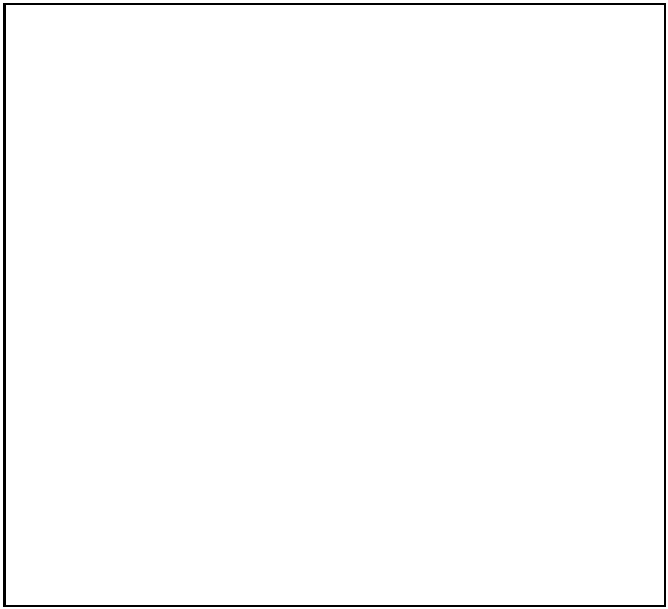
{The proton density profile n$_p$ for kT=7 keV. The position of the cooling
radius r$_c$ is marked. For kT=3 keV the density is lower by 0.04 dex.}
 \endfig

The last column in Table 1 gives the estimate of the gas mass
within 0.5 Mpc, which is about
1.2$\times$10$^{13}$ M$_\odot$: in the L$_{bol}$-Mass plane
(Edge and Stewart 1991a, their fig.7) this cluster falls close to the average
cluster relation.

The total gravitational mass of the cluster within 0.5~\-Mpc, computed
under the assumption of hydrostatic equilibrium (Fabricant et al. 1984),
is 1.8$\times$10$^{14}$ M$_\odot$ fot kT=7 keV and 7.5$\times$10$^{13}$
M$_\odot$ for kT=3 keV

\titlea{ Discussion}

The ROSAT observation described in this paper demonstrates that the inner
regions of the distant cluster MS0839.9+2938 are dominated by a cooling flow,
and that, very much like what is observed in two relatively nearby clusters
(Sarazin et al. 1992a, b), the space distribution of the cooling process
is highly inhomogeneous.

Here we would like to address the question whether photoelectric absorption
by cool gas within the cooling flow might be affecting the observed surface
brightness distribution. White et al. (1992) present spectral evidence, based
on EINSTEIN Observatory SSS data, that cold, X-ray absorbing gas
is often present in cooling flows, with an equivalent N$_H$ up to a few times
10$^{21}$ cm$^{-2}$: such a column density would significantly affect the
ROSAT HRI
counts. They argue on physical grounds that this gas is likely to be in
the form of dense and small clouds. If these clouds are regarded as remnants
of a cooling process which went on for 10$^{10}$ years, even if the space
distribution of their place of origin were highly inhomogeneous, it is
quite conceivable that their present,
cumulative distribution is a regular function
of the radial distance from the centre. Therefore the associated
absorption can hardly be responsible for the structures in the surface
brightness observed within the cooling radius. On the other hand, as it
was noted in Section 3, the brightness level at the barycentre of its
distribution, where the central giant elliptical is located, is lower
than the peaks in the surrounding structures. We suggest that
this situation could be due to photoelectric absorption. In fact,
a column density of 5$\times$10$^{21}$ cm$^{-2}$ would reduce the brightness
in the ROSAT
band by a factor of about 2 for kT $\geq$2 keV. A simple calculation shows
that,
if $\Dt M$ had remained constant over 10$^{10}$ years, and {\it all} the
cooled gas went into X-ray absorbing clouds, the N$_H$ that one obtains if the
clouds
were uniformly distributed within the cooling radius, and with a covering
factor of unity, is 7$\times$10$^{21}$ cm$^{-2}$. This value is a lower limit
to that which
would be obtained if the clouds were more realistically assumed to concentrate
towards the centre. We conclude that a spectral imaging of the cluster
will most probably reveal the presence of a strong absorption within 10" or
so from the centre, a goal achievable in the near future with the
angular resolution of the AXAF satellite.

\acknow{
We thank L. Norci for her generous help in a first
analysis of the ROSAT data at MPE and R. Della Ceca for his aid in the
astrometric calibration of the optical plate. Financial support
from GIFCO/CNR and from the italian Ministry for University
and Research is acknowledged.
}

\begref{References}

\ref
Arnaud K.A. 1988, in ``Cooling flows in clusters and galaxies",
Fabian A.C. (edt), Kluwer Academic Publ. Dordrecht, p.31

\ref
Cavaliere A., Fusco Femiano R., 1976, AA 49, 137

\ref
Edge A.C., Stewart G.C., 1991a, MNRAS 252, 414

\ref
Edge A.C., Stewart G.C., 1991b, MNRAS 252, 428

\ref
Fabricant D., Rybicki G., Gorenstein P., 1984, ApJ 286, 186

\ref
Gioia I.M., Maccacaro T., Morris S.L., et al.,
1987, in ``High redshift and primeval galaxies",
Bergeron J., Kunth D., Rocca-Volmerange B.,
J. Tran Thanh Van (eds), Paris, p. 231.

%\ref
%Gioia I.M., Maccacaro T., Morris S.L., Schild R.E., Stocke J.T., Wolter A.,
%1987, in ``High redshift and primeval galaxies", Proceedings of the 3rd
%IAP Astrophysics meeting, eds. J. Bergeron, D. Kunth, B. Rocca-Volmerange,
%J. Tran Thanh Van, Paris, page 231.

%\ref
%Gioia I.M. Maccacaro T., Schild R.E., Wolter A., Stocke J.T., Morris S.L.,
%Henry J.P. 1990, ApJS 72, 567

\ref
Gioia I.M. Maccacaro T., Schild R.E., et al., 1990, ApJS 72, 567

\ref
Jones C., Forman W., 1984, ApJ 276, 38

%\ref [TOGLIERE?]
%Jones C., Forman W., 1991, in 'Clusters and Superclusters of Galaxies', A.C.
%Fabian edt., NATO ASI Series C, pag.49

\ref
Mewe R., Gronenschild E.H.B., Oord G.H.J. van den, 1985 AAS 62, 197

\ref
MPE, 1991, ROSAT AO2 Call for Proposal, Max-Planck Institute fur
Extraterrestrische Physik, Garching b. Munchen.

\ref
Nesci R., Gioia I.M., Maccacaro T., et al., 1989, ApJ 344, 104

\ref
Sarazin C.L., O'Connell R.W., McNamara B.R., 1992a, ApJ 389, L59

\ref
Sarazin C.L., O'Connell R.W., McNamara B.R., 1992b, ApJ 397, L31

\ref
White D.A., Fabian A.C., Johnstone R.M., Mushotzky R.F., Arnaud K.A.,
1992, MNRAS 252, 72
\endref
\bye